\documentclass[graybox, envcountchap]{svmult}

\usepackage{mathptmx}        
\usepackage{amsmath}
\usepackage{amssymb}
\usepackage{color}
\usepackage{helvet}          
\usepackage{courier}         
\usepackage{dirtree}

\usepackage{makeidx}        
\usepackage{graphicx}        
\usepackage{subfig}

\usepackage{multicol}        
\usepackage[bottom]{footmisc}

\usepackage{hyperref}        
\hypersetup{colorlinks=true,urlcolor=blue}

\usepackage[misc]{ifsym}

\makeindex             

\def\prt{\partial}
\def\pt#1{\phantom{#1}}
\newcommand{\beq}{\begin{equation}}
\newcommand{\eeq}{\end{equation}}
\newcommand{\bal}{\begin{aligned}}
\newcommand{\eal}{\end{aligned}}
\newcommand{\rf}[1]{(\ref{#1})}

\def\fr#1#2{{{#1} \over {#2}}}
\def\half{{\textstyle{1\over 2}}}

\def\frac#1#2{{\textstyle{{#1}\over {#2}}}}
\def\ol#1{\overline{#1}}

\def\al{\alpha}
\def\be{\beta}
\def\ga{\gamma}
\def\de{\delta}
\def\ep{\epsilon}

\def\et{\eta}

\def\ka{\kappa}
\def\la{\lambda}
\def\rh{\rho}

\def\si{\sigma}

\def\ta{\tau}

\def\ph{\phi}
\def\ch{\chi}
\def\ps{\psi}
\def\om{\omega}
\def\Ga{\Gamma}
\def\De{\Delta}

\def\Om{\Omega}

\def\mn{{\mu\nu}}
\def\cL{{\cal L}}
\def\cG{{\cal G}}

\def\kb{{\overline k}}
\def\sb{{\overline s}}
\def\ab{{\overline a}}
\def\cb{{\overline c}}

\def\kf{\hat k_F}
\def\kaf{\hat k_{AF}}

\def\kde{\hat\ka_{DE}}
\def\kdb{\hat\ka_{DB}}
\def\khe{\hat\ka_{HE}}
\def\khb{\hat\ka_{HB}}

\def\voc{\mathrel{\rlap{\lower0pt\hbox{\hskip1pt{$c$}}}
    \raise3pt\hbox{$\neg$}}}
\def\vok{\mathrel{\rlap{\lower0pt\hbox{\hskip1pt{$k$}}}
    \raise6pt\hbox{$\neg$}}}

\def\ol#1{\overline{#1}}
\def\a{$a_\mu$}
\def\b{$b_\mu$}
\def\c{$c_{\mu\nu}$}
\def\d{$d_{\mu\nu}$}
\def\e{$e_\mu$}
\def\f{$f_\mu$}
\def\g{$g_{\la\mu\nu}$}
\def\H{$H_{\mu\nu}$}

\begin{document}


\title{Testing Gravity in the Laboratory}

\author{Quentin G.\ Bailey}

\institute{ (\Letter) \at Embry-Riddle Aeronautical University, 3700 Willow Creek Road, Prescott, Arizona, USA \email{baileyq@erau.edu}
}

\maketitle

\abstract{In this chapter we discuss recent work on precision Earth laboratory tests of different aspects of gravity.
In particular the discussion is focused on those tests that can be used to probe hypothesis for physics beyond Newtonian gravity and General Relativity. 
The latter includes tests of foundations like local Lorentz invariance, 
Weak Equivalence Principle tests, 
short-range gravity tests, 
gravimeter-type tests, 
and other frontier possibilities like the free-fall of anti-matter and searches for non-Riemann gravity effects.
The focus is on key results in theory, 
phenomenology, 
and experiment in the last few decades.
We describe the motivations for continued interest in precision tests of gravity in the laboratory, 
including the possibility to search for physics beyond General Relativity.
Test frameworks for describing deviations from General Relativity are emphasized, 
including ones based on effective field theory, 
including allowing for generic violations of Lorentz symmetry, CPT symmetry, 
and diffeomorphism symmetry.
}


\section{Early history and motivations for testing gravity with precision laboratory instruments}
\label{earlyhistory}

As laboratory experiments evolved during the renaissance to the current epoch, 
a number of early experiments concerning the nature of gravity on Earth were conceived.  
These early tests established some of the most important basic principles of the currently accepted highly successful theory of General Relativity (GR).

Early work by Galileo established the universality of freefall for objects of different mass (and composition) using simple inclined planes (establishing that all objects fall with the same local acceleration on the Earth's surface when air resistance can be neglected).
Later, 
Newton's universal law of gravitation confirmed this property as part of the nature of the gravitational force depending on the product of the masses that attract one another, and the inverse square distance, 
often dubbed the inverse-square law (ISL):
\beq
F_{\rm grav} = \fr {G m_1 m_2}{r^2}, 
\label{isl}
\eeq
where $G$ is Newton's gravitational constant, 
$m_1$ and $m_2$ are the two masses and $r$ is the distance between the masses.  
The force points along the line between the masses, 
and is consistent with Newton's third law 
(so that the force of attraction on mass $1$ is equal 
and opposite to the force of attraction on mass $2$).

As an ingeneous application of a torsion pendulum apparatus, 
E\"otv\"os studied the effect of the Earth's gravity on laboratory sized masses of differing composition \cite{Eotvos1890}. 
The basic torsion balance consisted
of a light horizontal beam suspended by a metal wire.  On one end of the beam, 
was suspended a mass, 
on the other end, 
another mass with differing composition.
(See figures 1,2 in \cite{Bod:1991jb} and the figure in \cite{Dicke1961}).
The null result otained for the relative acceleration on Earth's surface of two species with the same mass,
confirmed to a high degree of accuracy the Weak Equivalence Principle (WEP).
This principle states that all bodies fall in a gravitational field with the same acceleration.
The experiment was improved upon in the intervening years to the modern era
{e.g., \cite{Dicke1961}}.

The gravitational redshift effect of GR predicts that in the vicinity of a large mass, 
like being on Earth's surface, 
clocks will run at slower rates than clocks farther from the Earth.
For example a measured frequency shift for emitted light between two locations in 
a static gravitational field is given by $\de \nu/\nu = \De U$, 
where $U$ is the gravitational potential.
In fact, 
near Earth's surface a change in ticking rates of clocks 
can be detected in sensitive tests (this is used routinely 
in GPS satellites for precision navigation \cite{Ashby:2003vja}).
In the 20th century some early tests of the gravitational redshift were key in verifying foundations of GR
\cite{Will:2018bme}.
For instance, 
the classic Pound-Rebka-Snider experiments in the early 1960s measured frequency shifts from ascending and descending atomic emissions (gamma ray photons) using a 25-m tower.
\cite{Pound60,Pound65}.
The 1965 measurement verified the redshift effect to about the $1\%$ level.
Further improvements in the modern era in redshift tests and motivations for violations
of the GR prediction are discussed in the next sections.

The precise measurement of the gravitational constant $G$ represents 
another history of successive improvements, 
though not as striking as other measurements of fundamental constants \cite{Chao20}.
Many applications of Newton's law of gravitation involve orbital mechanics and observations.
In these scenarios, 
in the the observables 
that occur, 
the constant $G$ is always paired with a mass as in $GM$.
For instance, 
one can see this from considering elliptical orbits around a central body $M$.
The time for one orbit $T$ is given by Kepler's third law, $T = \fr {2\pi a^{3/2}}{\sqrt{GM}}$.
To extract a value for $G$,
one needs to separate $G$ from the values of the masses involved, 
a task most readily achieved by measuring the force of gravity 
in a laboratory setting \cite{Chao20}.
Notable early experiments measuring $G$
include the classic torsion balance experiment by 
Cavendish \cite{Cavendish}.

Foundations of General Relativity include not only the WEP, 
but also local Lorentz invariance, 
part of the Einstein Equivalence Principle.
Local Lorentz invariance holds that experimental results in a freely falling inertial frame are independent or the orientation and the velocity of the laboratory.

Local Lorentz invariance was tested historically with the Michelson-Morley experiment,
which tested the assumption that electromagnetic waves propagate through a medium, 
obtaining the classic null result \cite{mm1887}.
While the Michelson-Morley experiment established the isotropy of the speed of light, 
later tests, 
like
the Kennedy-Thorndike experiment in the 1930s \cite{kennedy32}, 
looked for dependence
on the laboratory velocity.  
With these tests and many others,
local Lorentz invariance has become the standard tested assumption for modern physics. 
However, 
we discuss below continued motivations for a plethora of ever more precise tests occurring in recent decades.

\section{Laboratory tests in the Modern era}
\label{ModernEra}

Physicists worldwide have continued to test the inverse square law \rf{isl}, 
as well as search for laboratory-based signatures of General Relativity, 
and proposals for physics beyond GR.
In addition, 
modern precision metrology and Geophysics have played a role in the development of sensitive instruments for mapping the gravitational field of the Earth, like gravimeters.
These instruments can also be used to test gravity as discussed below.

Tests of WEP on Earth and in near Earth orbit have been steadily improved over the last decades, 
as arguments about what physics beyond GR might look like have pointed to possible WEP violations.
Gravitational redshift tests have also been of interest and undergone improvement.
Local Lorentz invariance continues to be tested, 
experiencing a recent boom in tests due to the motivation for searches for new physics.
Modern tests use all manner of precision instruments like resonant cavities and atomic clocks.

In principle, 
modern laboratory precision experiments {\it on Earth} include gravitational wave observatories
\cite{GWdisc16}, 
albeit the focus there is on astrophysics.
However, 
the discussion of these tests is contained in several other chapters in this book, 
so we do not go into details here.

\subsection*{Motivations}
\label{motivations}

A key reason for the continued interest in developing ever more precise instruments for testing gravity on Earth or in near Earth orbit comes partly from the developments in theoretical physics in the 20th and 21st century.
We describe some major areas below where theoretical ideas have pushed for new sensitive tests.
This discussion naturally includes tests of founding principles of GR, 
like local Lorentz invariance, 
the latter topic with a long history and recent boom in activity.

\subsubsection{Tests of Local Lorentz invariance}
\label{Tests of Local Lorentz invariance}

In the mid 20th century, 
phenomenological test models for Special Relativity (SR), 
describing small deviations from SR, 
were developed using the idea of modified Lorentz transformations.
These modified transformations describe how to go from a special reference frame where light propagation is isotropic to a ``moving frame” where anisotropy could be potentially detected.  
An early version was countenanced by Robertson \cite{robertson49}, 
while Mansouri and Sexl added to this early work in 1977 \cite{RMS77}.
The result was a three parameter model, 
the ``RMS" test model,
with parameters $\al$, $\be$, and $\ga$ describing small deviations from the usual Lorentz transformation. 
Special Relativity holds when these parameters vanish.  
While the RMS test model was a bellweather for several decades, 
it was more recently shown that it has inconsistencies when compared to field theory approaches \cite{km13}.

Given the many tests of the 20th century, 
the main motivation for continued interest in precision tests of Lorentz symmetry has come from a number of theoretical works.
For example, 
Kosteleck\'y and Samuel showed that the vacuum of string field theory might break Lorentz symmetry spontaneously \cite{ksstring89}.
Other ideas published in the literature suggest the possibility of spacetime symmetry breaking through a variety of mechanisms.
For example, 
quantum gravity models \cite{gp99},
non-commutative field theories \cite{Mocioiu:2000ip,chkl01}, 
and extra dimensions \cite{Overduin:2021wdq}
can lead to observable Lorentz violation.
More comprehensive reviews of the literature on
mechanisms for spacetime-symmetry breaking can be found in Refs.\ \cite{Mariz:2022oib,Will:2014kxa,Mattingly:2005re,Addazi:2021xuf,Tasson:2022obv,Hees:2016lyw}.

Some other recent approaches for frameworks testing Lorentz invariance
include modified dispersion relations.
This is motivated by approaches to quantum gravity \cite{camelia01}.
The dispersion relations typically take the form, 
for a particle with mass $m$, 
of $E^2-p^2-m^2 - {\tilde L}_p p^2 E+...=0$, 
where ${\tilde L}_p$ is of order of the Planck length.
These and other ``kinematic" test equations have been used in astrophysics to constrain the extra terms, 
like measurements of high energy photons, 
but care is required with interpretations \cite{Jafari:2006rr,km09}.
For recent reviews of this type of analysis see \cite{Addazi:2021xuf,Liberati13,Mattingly:2005re}.

In the late 90s and early 2000s, 
a new approach, 
involving the systematic development of a fully dynamical effective field theory (EFT) framework for testing relativity principles arose in a number of key publications \cite{ck97,ck98,km02,k04}. 
The basic premise is first that local Lorentz symmetry holds to a high approximation, 
so any deviations are small.  
Such deviations then must be describable by the coupling of a ``matter” field to a generic background tensor field with components called ``coefficients for Lorentz violation”.  
An EFT framework that includes GR and the Standard Model of particle physics, 
can allow one
to systematically categorize all possible terms 
that break symmetries like CPT and Lorentz symmetry.
This description, 
as it turns out, 
can be matched to many 
proposed models for quantum gravity and other ideas as well as test frameworks.
Many tests have now been analyzed within the framework \cite{km09, km13, datatables}, 
as it allows cross comparisons between vastly different tests (e.g., astrophysics versus Earth-laboratory tests).

The systematic description offered by the EFT framework includes possible breaking of all
spacetime symmetries.
This includes local Lorentz symmetry,
the discrete CPT symmetry, 
and diffeomorphism symmetry (the latter relevant for gravity and curved spacetime).
The EFT framework allows comparison of measurements across all possible precision tests with matter, light, gravity, and others.

Various observables for laboratory tests have been studied in this framework.
This includes short-range gravity tests \cite{bkx15,Shao:2015gua,Shao:2016cjk}, 
gravimeter measurements \cite{Muller:2007es,Flowers:2016ctv},
WEP tests \cite{Bars:2019lek}, 
and redshift tests \cite{Hohensee:2011wt}.
With the EFT framework, 
systematic tests of local Lorentz invariance can be achieved in any conceivable experiment.  
Results from any experiments probing the same ``sector” can be compared for their measurements of the coefficients.  
For instance, 
results from cosmic rays can be compared with results from table-top tests with matter as described below.
 
Consider the electromagnetic sector of the EFT framework.
This effectively contains symmetry-breaking corrections to the Maxwell equations \cite{km02}.
The basic Lagrange density for this sector assumed $U(1)$ gauge invariance but breaks global Lorentz symmetry and CPT symmetry \cite{km09}:
\beq
{\cal L} = - \frac 14 F^\mn F_\mn - \frac 12 F_\mn ({\hat k}_F)^{\mu\nu\ka\la} F_{\ka\la} 
+\frac 12 \ep^{\ka\la\mu\nu} A_\la ({\hat k}_{AF})_\ka F_\mn
\label{photon}
\eeq
The symmetry breaking occurs when the system under study rotates or is boosted with respect to the background coefficients and the coefficients are zero for perfect spacetime symmetry (Lorentz and CPT).
The framework in \rf{photon} yields the following modified Maxwell equations in source-free space:
\beq
\bal
{\bf \nabla}\cdot {\bf D} & = 0,
\notag\\
{\bf \nabla} \times {\bf H} - \prt_0 {\bf D} & = 0 ,
\label{maxeq}
\eal
\eeq
where
\beq
\bal
{\bf D} &= {\bf E} +2 {\bf \kaf} \times {\bf A} 
+{\bf \kde} \cdot {\bf E} + {\bf \kdb} \cdot {\bf B}, 
\notag 
\\
{\bf H} &= {\bf B} -2(\kaf)_0 {\bf A} + 2 {\bf \kaf} A_0
+ {\bf \khb} \cdot {\bf B} 
+{\bf \khe} \cdot {\bf E}.
\notag\\
\eal
\eeq
The coefficient matrices appearing
are related to the underlying coefficients in the Lagrange density as follows:
\beq
\bal
(\kde)^{jk} &= -2(\kf)^{0j0k} , 
\notag\\
(\khb)^{jk} &= \half(\kf)^{lmrs} \ep^{jlm}\ep^{krs} , 
\label{kappas} \\
(\kdb)^{jk} &= -(\khe)^{kj} = (\kf)^{0jlm}\ep^{klm}. 
\eal
\eeq
Thus the effects of Lorentz and CPT violation in vacuum loosely resemble
behavior in anisotropic media.
The ``hats" on the coefficients stand for general operators consisting of a series of constant  coefficients coupled with derivatives.
In this sample discussion here we confine our attention to the lowest order constant coefficients.
Thus, 
$(k_F)^{\ka\la\mu\nu}$ contain $19$ coefficients dimensionless coefficients, 
while there are $4$ $(k_{AF})_\ka$ coefficients with mass dimension $M$.

Numerous measureable modifications to electromagnetic tests can be calculated from the results above.
This includes shifts in resonant cavity frequencies \cite{Mewes:2012sm}, 
birefringence and dispersion of photons \cite{km01,km02}, 
modifications to electric and magnetic fields in the laboratory
\cite{bk04}, 
and many others \cite{Yoder:2012ks,Tobar:2009gw,Frank:2006ww}.
Current limits from some of the latest experiments and observations are discussed below in section \ref{recentexp}.

Modifications of the matter sector have also been considered. 
Another place where new physics, 
in the form of spacetime symmetry breaking observables, 
may arise, 
is the Fermion sector of the EFT.
The description is written in terms of a basic complex-valued spinor field $\ps$, 
representing an electron, proton, neutron or other fermion.
The Lagrange density of the framework takes the form
\beq
\label{Lps}
\cL_\ps =  \frac {1}{2} i \ol{\ps} \Ga_\nu D^\nu \ps 
- \ol{\ps} M \ps,
\eeq
where
\beq
\Ga_\nu \equiv \ga_\nu + c_\mn \ga^\mu + d_\mn \ga_5 \ga^\mu 
   + e_\nu + i f_\nu \ga_5
   + \half g_{\la \mu \nu} \si^{\la \mu},
\eeq
and
\beq
M \equiv m + a_\mu \ga^\mu + b_\mu \ga_5 \ga^\mu 
   + \half H_\mn \si^\mn.
\eeq
Here \a, \b, \c, \d, \e, \f, \g, and \H\
are fermion-sector coefficients for Lorentz violation,
$\ps$ is the fermion field,
and the $\ga^\mu$ are the usual Dirac matrices.
Note that $D_\mu = \prt_\mu + i q A_\mu$,
is the covariant derivative in this context, 
with the vector potential $A_\mu$. 
Setting the coefficients to zero
results in the Lorentz-invariant limit
reproducing the conventional Dirac lagrangian.

One can calculate the effect of the extra terms,
using a Hamiltonian description, 
on energy levels in atoms, 
and thereby construct experimental signals that can be measured by sensitive clocks \cite{kl99}.
For example, 
relevant for atomic clocks the 
formula for the extra piece of the hamiltonian can be written as
\beq
\de h = -\frac {1}{m} c_{jk} p^j p^k + \frac {1}{2m^2} b_j \si^j p^2 +...,
\label{atomicshift}
\eeq
which displays the nonrelativistic contributions from the general case \rf{Lps}.
The ellipses includes contribution from the other coefficients.
This result has been generalized to include additional ``nonminimal" parts of the EFT \cite{kv15,kv18}.
Clocks tests are discussed further in the experiment section.

\subsubsection{WEP tests, theory}

There remains wide interest in precision tests of WEP in the modern era.
Firstly, sensitivities have vastly improved and measurements in space are now possible.
Recent results have confirmed the universality of freefall to a very high degree of precision for different types of matter.
The basic figure of merit often adopted is the E\"otv\"os parameter $\eta$ given by
\beq
\et = \fr {2 |a_1 - a_2|}{|a_1+a_2|},
\label{}
\eeq
where $a_1$ and $a_2$ are the accelerations of bodies $1$ and $2$.
Of course, when WEP holds $\et=0$.

Theoretical ideas abound that could produce a nonzero $\et$ for bodies with differing composition.
For example, 
the authors in Ref.\ \cite{fischbach86} showed that an extra gravitational potential between two bodies could 
arise from exchange forces 
with vector of scalar bosons from an unknown interaction.
This would take the form
$\de V \sim q_1 q_2 \exp (-r/\la)$,
where $q_1$ and $q_2$ are the charges of each body associated with the new hypothetical force, 
$r$ is the distance, 
and $\la$ is the length scale of the interaction.
Other examples include WEP violations arising from string models \cite{Damour:1994zq},
chameleon fields
\cite{Khoury04}, 
and more (see a summary in Ref.\ \cite{Will:2018bme}).

One particularly interesting subset of the EFT framework, 
are the matter-gravity coupling terms.
These would be detectable in WEP-type tests, 
or tests with antimatter, 
where the freefall acceleration for two distinct types of matter
is compared. 
Kosteleck\'y and Tasson in Refs.\ \cite{kt09,kt11} use the EFT framework to study
the effects of matter gravity couplings.
The underlying results are based on a generalization of the matter Lagrange density 
\rf{Lps} to allow for gravitational interactions.
With spinors and gamma matrices present, 
one uses the vierbein formalism to incorporate them with gravity
\cite{Weinberg:1972kfs,Hehl:1976kj,k04}.

In an effective classical limit, 
where spin is averaged out, 
one obtains the following effective action for
a body $B$, 
which could be composed of several species of particles:
\beq
S^B = \int d \la \big[   -m^B \sqrt{-(g_\mn +2 (c^B)_\mn) \fr {dx^\mu}{dx^\la}{dx^\nu}{dx^\la} } 
-(a^B)_\mu \fr {dx^\mu}{d\la} \big].
\label{Blag}
\eeq
Here there are $10$ coefficients for Lorentz violation contained in
$(c^B)_\mn$ and there are $4$ in $(a^B)_\mu$.
Using a suitable model of a body $B$ one can write these coefficients in terms of coefficients for constituent species $w$.
Usual, 
with ``ordinary matter" this includes $w=e,p,n$, for electron, proton, neutron.
This increases the number of coefficients by a factor of $3$, 
and different bodies can have differing dependencies on the constituent particle level coefficients \cite{kt11}.

Using this action \rf{Blag}, 
one can calculate observables for a number of distinct tests. 
For example, 
the following modified Newton's second law equation holds for masses near the surface of a planet:
\beq
F_{\hat j} = m_{{\hat j}{\hat k}} a^{\hat k}, 
\label{modforce}
\eeq
where $a^{\hat k}$ are the components of the laboratory measured acceleration
and $m_{{\hat j}{\hat k}}$ is the inertial mass matrix.
For example, 
$F_{\hat j}$ has components in all $3$ laboratory direction, 
\beq
F_{\hat z} = -g [ m^T + 2\al \ab^T_{\hat t} + 2\al \ab^S_{\hat t} + (c^T)_{{\hat t}{\hat t}} 
+ (c^S)_{{\hat t}{\hat t}} ].
\label{Fz}
\eeq
In this equation the coefficients $\ab_\mu$ and $c_\mn$ 
are evaluated on the local time $\hat t$ direction c, 
with superscripts $T$ and $S$ standing for the test and source body, 
respectively.
The constant $\al$ is a model-dependent constant that depends 
on the coupling of gravity to matter.
Meanwhile the inertial mass matrix is
\beq
m_{{\hat j}{\hat k}} = m^T ( 1 + (c^T)_{{\hat t}{\hat t}} ) + 2 m^T (c^T)_{{\hat j}{\hat k}}.
\label{mjk}
\eeq

The results \rf{modforce}-\rf{mjk} displays several interesting features.  
First, 
components of the coefficients $\ab_\mu$ and $c_\mn$ show up in the expressions in the Earth lab frame.
However, 
assuming constancy of the coefficients in the standard Sun-centered Celestial Equatorial frame (SCF) (see figure \ref{scf} and discussion in \ref{local Lorentz invariance tests}), 
the signal actually acquires a specific sidereal time dependent signature \cite{kl99}. 
Furthermore, 
the result indicates the possibility of differing rates of fall for species of matter, 
thereby also breaking the WEP.

We mention in passing the WEP principle can applied to
bodies bound by gravitation and internal forces,
which is the so-called GWEP or Gravitational Weak equivalence principle as part of the Strong Equivalance Principle
\cite{Will:2018bme}.
General Relativity obeys this principle while alternatives can break it.
We do not discuss this in detail since tests of GWEP are typically beyond the Earth laboratory but the reader is referred to the literature \cite{Nordtvedt68,Voisin:2020lqi,Shao:2019cyt}.

\subsubsection{Non-Newton forces}
\label{non-Newton}

Corrections to Newton's laws, 
distinct from those arising in GR, 
have been proposed based on explorations
of candidates for unified theories of physics.
One notable idea came from a paper on the the hypothetical Dilaton particle coupling to gravity
\cite{Fujii:1971vv}.
The dilation is the hypothetical Nambu-Goldstone boson associated with dilatation invariance \cite{Carruthers:1971vz}.
The coupling with gravity was shown to produce a modification of the gravitational force at short distances.
This,
in part, 
led to ``parameterizations" of deviations from 
Newton's inverse square law \cite{Adelberger:2009zz,Murata:2014nra}.

The idea is that gravity could be modified by some new interaction (a new force for instance) that would only show up on short distance scales (e.g., submilimeter).
The interaction resembled the Yukawa force
from Quantum Field Theory for a massive particle mediator.
A commonly used version is written in terms of a modified gravitational potential energy between two masses
\beq
V_{\rm Yukawa} = -\fr {G m_1 m_2}{r} 
\left( 1+ \al e^{-r/\la} \right), 
\label{yuk}
\eeq
where $\al$ and $\la$ are parameters to be measured controlling the strength and the length scale of the new interaction, respectively.

Many proposed models of physics beyond the Standard Model and GR include ``force corrections" that take the form \rf{yuk}.
The parameters $\al$ and $\la$ can be mapped to various constants and parameters in the proposed models.
For example, 
one key proposal was a ``fifth force" hypothesis wherein the extra $\al e^{-r\la}$ arises from an intermediate range coupling between the two masses
that could depend on baryon number or hypercharge
\cite{fischbach86}.

Another phenomenological framework allows for variation in the $1/r^2$ falloff of the Newtonian force.
For instance, 
a power law expansion has been used:
\beq
V_{\rm power}= -\fr {G m_1 m_2}{r} \left( 
1+ \sum_k \fr {\be_k}{r^k} \right),
\label{power}
\eeq
with the various length dimension parameters $\be_k$ describing the strength of the new power-law force.
This test model is motivated by proposals for extra dimensions of space.  
For instance, 
the proposal of Ref.\ \cite{Arkani-Hamed:1998jmv},
involve the introduction of two extra space dimensions on small length scales, 
which can thereby resolve the hierarchy problem in particle physics.

More recently, 
using a general test framework described in section \ref{Tests of Local Lorentz invariance}, 
it has been shown that short-range tests could be of interest for probing 
certain kinds of Lorentz violation.
In the gravity sector of this framework there are terms beyond GR that include possible Lorentz violation and diffeomorphism violation.
The Lagrange density is a series taking the form $\cL = \cL_{GR} + \cL^{(4)} + \cL^{(5)}+ \cL^{(6)}+...$, 
wwith each subsequent term containing couplings of background tensors $t_{\mu\nu\la...}$ with successively higher powers of spacetime curvature $R_{\mu\nu\ka\la}$ and covariant derivatives
$\nabla_\mu$ \cite{bkx15}.
The superscript label indicates the mass dimension $M^d$ of each term.
For example, 
the leading symmetry breaking corrections from this framework can be evaluated for weak-field gravity.
They take the form of a Lagrange density written in terms of 
the metric fluctuation $h_\mn$ around a flat background \cite{bk06,km16,bh17}.
This series resembles
\beq
\cL \supset \frac {1}{4\ka} \sb^{\mu\ka} h^{\nu\la} \cG_{\mu\nu\ka\la} 
-\frac{1}{16\ka} h_{\mu\nu} (q^{(5)})^{\mu\rh\al\nu\be\si\ga} \prt_\be R_{\rh\al\si\ga}+...,
\label{gravL}
\eeq
with coefficients $\sb^\mn$ and $(q^{(5)})^{\mu\rh\al\nu\be\si\ga}$, 
and the constant $\ka=4\pi G_N$.
Here, 
$\cG_{\mu\nu\ka\la}$ is the linearized double-dual curvature tensor.

For short-range gravity tests, 
the effects of the mass dimension $6$ terms can be examined
in the limit of weak fields.
This yields a Newtonian potential energy between two point masses receives a correction $\de V_{LV}$ given by
\beq
V_{\rm LV} (\vec{r}) 
= - G_N \fr{{m_1}{m_2}}{|\vec{r}|^{3}}{\ol k}(\hat{r},T) \left( \frac 32  (\kb_{\rm eff})_{jkjk}
- 9  (\kb_{\rm eff})_{jkll} \hat{r}^j \hat{r}^k 
+ \frac {15}{2} (\kb_{\rm eff})_{jklm} 
\hat{r}^j \hat{r}^k \hat{r}^l \hat{r}^m \right),
\label{pot}
\eeq
where $\vec{r}=\vec{x}_{1}-\vec{x}_{2}$ is the separation vector for the point masses $m_1$ and $m_2$.
The coefficients for symmetry-breaking $(\kb_{\rm eff})_{jklm}$ have length squared dimension
and the combination occurring has 14 {\it a priori} independent coefficients (the rotation invariant $(\kb_{\rm eff})_{jkjk}$ vanishes.  
There is also a generalization of \rf{pot} for coefficients of higher-mass dimension terms in the Lagrange density (e.g., higher powers of curvature) \cite{km17}.
For this particular result \rf{pot}, 
a perturbative method has been used where it is assumed that $V_{LV}$ is small compared to the Newtonian potential, 
or equivalently, 
the accompanying force is smaller than the Newtonian force.  
This means experiments that can measure the nominal Newtonian force are more interesting probes of \rf{pot}, 
these are the ones in the lower right of the plot \rf{alphalambda}.

In addition to the result \rf{pot}, 
recently it has been proposed that
spacetime symmetry breaking could indeed lead to large forces on small length scales and still be consistent with existing constraints \cite{Bailey_2023}.
For example, 
if one restricts attention to isotropic coefficients in $\cL^{(6)}$, 
the exact modification to the Newtonian potential can be calculated in terms of $3$ combinations of coefficients in the action labelled $k_1$, $k_2$, and $k_3$.
One particular special case takes the form,
\beq
V_{LV}= 
- G_N \fr{{m_1}{m_2}}{2|\vec{r}|} 
\exp \left( - \fr {r}{\la_1} \right)
\Bigg[ \cos \left( \frac {r}{\la_2} \right)
\mp \frac {(1+2 \ch)}{\sqrt{|1+4\ch|}} 
\sin \left( \frac {r}{\la_2} \right),
\Bigg]
\label{pot2}
\eeq
where $\ch=(k_2+k_3)/k_1$.
The second term in this expression shows that an arbitrarily large non-Newtonian force could arise.

An interesting point about signals in \rf{pot} and \rf{pot2} is that many of the same coefficients for symmetry breaking can also be probed in 
astrophysical tests, 
like gravitational waves.
For example, 
in terms of coefficients in the underlying action, 
$k_2 = ( s^{(6)0kl0klmm} + k^{(6)0k0klmlm})/15$.
In analysis of gravitational waves, 
one finds an isotropic coefficient $(k_{I}^{(6)})_{00}$
controlling symmetry-breaking dispersion effects \cite{km16,Mewes:2019}.
This coefficient can be expressed in terms of coefficients in the action as
$(k_{I}^{(6)})_{00} =\sqrt{4\pi} (
s^{(6)0ij0ijkk} /5
+k^{(6)0i0ijkjk}/15 +...)$ \cite{Bailey_2023}.
So there is an overlap of coefficients controlling the effects in short-range gravity to those in gravitational waves.
The EFT thus allows comparison using tests
``across the universe" hunting for the same quantities \cite{shao20,Gong:2023ffb,datatables}.

\subsubsection{Gravitational Redshift}

Interest in the gravitational redshift tests has continued in the last decades.
The commonly used test equation inserts a parameter in front of the GR redshift term in the expression for the frequency shift of an electromagnetic wave:
\beq
\fr {\de \nu}{\nu} = (1+\al ) \De U,
\label{redshift}
\eeq
where $\De U$ is the difference in the gravitational potential difference between two locations where the frequency difference $\de \nu$ is calculated.
Models beyond GR that produce a nonzero value for $\al$ involve assuming that the ticking rates of clocks may vary with position (a violation of ``Local Position Invariance") \cite{Will:2018bme}.
For example, 
consider using a modified action for charges and electromagnetic fields called the ``TH$\mu\ep$"
formalism. 
It has been shown that, 
for a range of values of the parameters in the model, 
LPI is broken \cite{will74}.

Considering the EFT described above, 
redshift tests are sensitive to different forms of Lorentz violation mentioned.
In Ref.\ \cite{kt11}, 
the description of the modification of gravity and matter due to Lorentz and CPT violation yields modifications to the GR redshift prediction.
The ratio of the frequency of electromagnetic waves between two positions (emission $e$ and reception $p$) in the gravitational field above the surface of a planet can be written,
\beq
\fr {\nu_p}{\nu_e} = G_N m^S \fr {r_e-r_p}{r_e r_p} \left( 1+ \fr {2 \al}{m^S} (\ab^T)_0 + (\cb^S)_{00} + \xi_{\rm clock}+...\right),
\label{SMErs}
\eeq
where the ellipses include contributions from the $\sb_\mn$ coefficients.
The term $\xi_{\rm clock}$ depends on the coefficients in the matter sector and varies for distinct types of clocks.
A separate calculation is required to determine the value of $\xi_{\rm clock}$ \cite{Hohensee:2011wt}.
For instance, 
for a clock based on the Bohr levels of hydrogen, 
$\xi_{\rm Bohr} \approx -2(m^p (\cb^e)_{00} + m^e (\cb^p)_{00})/(3m^p)$.
Typically the value of $G_N m^S$ must also be calibrated, for example from orbital measurements.
This can sometimes cancel the signal for the redshift \cite{Hohensee:2010cd}.
It turns out, 
however, 
for the expression \rf{SMErs}, 
the combinations of coefficients in orbital tests signal are distinct so one can perform a ``calibration" and still have a measureable signal distinct from GR.

\subsubsection{Variations in local value of g}

As precise measurements of the Earth's gravity field and the Earth tides have improved over the 20th century, 
it became apparent that predictions of GR or modified models might be tested with precision measurements on Earth, 
such as with state-of-the-art gravimeters.

It was proposed in the 1970s to look for modifications to the local value of $G_N$ with the Parametrized Post-Newtonian (PPN) framework \cite{Nordtvedt:1972zz}. 
The signal was written as a variation
in the Newtonian constant $G_N$ that could be measured in a laboratory:
\beq
\fr {\De G_N}{G_N} = -\frac 12 ( \al_1 -\al_3 -\al_2 )]w^2 
-\frac 12 \al_2 (\vec w \cdot {\hat n} )^2 +..., 
\label{deltaGPPN}
\eeq
where the PPN parameters $\al_1, \al_2, \al_3$ control the degree of Lorentz violation in the metric-based framework \cite{Will:2018bme}.
With isotropy of PPN parameters assumed in a preferred frame like the CMB rest frame, 
$\vec w$ is the velocity of the laboratory through the CMB and $\hat n$ is the unit vector from the test mass to the Earth.

More recently it was shown in Ref. \cite{bk06}, 
that in a similar manner to the PPN above, 
the variation in the local value of $g$ could measure some of the coefficients gravity sector of the EFT framework for testing spacetime symmetries.
For instance,
in addition to (large) Newtonian tidal effects,
Lorentz-symmetry breaking terms in \rf{gravL} result 
in extra variations in the locally measured value of $g$:
\beq
\fr {\de g}{g} = 1 + \frac 32 \sb^{TT} 
+ \frac 12 \sb^{\hat z \hat z} 
- \sb^{T \hat z} V_{\oplus}^{\hat z}
-  3 \sb^{TJ} V_{\oplus}^J+...,  
\label{delg}
\eeq
where the coefficients $\sb_\mn$ are projected along laboratory frame basis ${\bf e}_{\hat j}$, 
Here, 
$T$ and $J$ refer to coordinates in the special frame described in \ref{local Lorentz invariance tests}, 
and $V_{\oplus}^J$ is the velocity of the Earth relative to the Sun.
Gravimeter tests can measure combinations of coefficients in $\sb_\mn$
using \rf{delg}, 
as detailed below in section \ref{recentexp}.

\subsubsection{Torsion and nonmetricity}

The idea that GR could be modified to include torsion or nonmetricity is of general interest in the gravity theory community \cite{Hehl:1976kj,Ni:2009fg}.
The interest in torsion is motivated by work with models of gravity including spin as a source.
It turns out certain experiments designed to test Lorentz symmetry for matter on Earth can be sensitive to the presence of spacetime Torsion and Nonmetricity.
Thus this is another way for a laboratory test to probe gravity indirectly.

The basic theory uses Riemann-Cartan spacetime or spacetime with nonmetricity.
For Riemann-Cartan spacetime, 
one defines the covariant derivative using
\beq
\bal
\nabla_\mu g_{\nu\la} &= 0,
\\
\nabla_\nu e_\mu^{\pt{\mu}a}=0,
\label{metricity}
\eal
\eeq
where $e_\mu^{\pt{\mu}a}$ is the vierbein, 
which can be used to incorporate spinor fields into the theory.
While the covariant derivative of a vector, 
for example, 
still takes the form
\beq
\nabla_\mu v^\la = \prt_\mu v^\la + \Ga^\la_{\pt{\la}\mu\al} v^\al, 
\label{covderiv}
\eeq
the connection can have an antisymmetric piece $\Ga^\la_{\pt{\la}[\mu\al]} = T^\la_{\pt{\la}\mu\al}/2 $, 
where $T^\la_{\pt{\la}\mu\al}$ is the torsion tensor.
On the other hand, 
broadening the usual ``metricity" assumption of \rf{metricity}
leads to the nonmetricity tensor 
$Q_{\mu\nu\la} = \nabla_\mu g_{\nu\la}$.

Using an action approach
with the veirbein to incorporate spinors, 
one can include the effects of torsion and nonmetricity for laboratory test particles.
For instance one can calculate the effects on trajectories and Hamiltonians relevant for lab searches. 
Limits have now been placed on components of the torsion tensor and the nonmetricity tensor in the laboratory \cite{Kostelecky:2007kx,Lehnert:2013jsa,Foster:2016uui}.
These results take advantage of the similarity between some torsion effects on matter, 
and the CPT and Lorentz-violating terms in the action \rf{Lps} 
using results from prior experiments using the EFT framework \cite{datatables}.

\subsection*{Recent laboratory experiments}
\label{recentexp}

\subsubsection{Local Lorentz Invariance tests}
\label{local Lorentz invariance tests}

To probe the various types of hypothetical Lorentz violation, 
modern day Michelson-Morley tests have been developed over the last decades with ever increasing sensitivity. 
These typically involve the use of high precision resonant cavities at locations around the globe.
Other precision tests with atomic clocks and other apparatus have been performed. 

In Table \ref{SRtests} we display a partial list of recent precision experiments focused on testing Lorentz invariance for differing matter or fields in the Earth laboratory setting.
Note also, these tests are complemented by a vast number of astrophysical tests on matter and photons (beyond the full scope of this article but see \cite{datatables} for a comprehensive list of recent tests and \cite{Will:2018bme} for a list of tests going back $100+$ years).  

\begin{table}
\caption{A partial list of experiments in the last decades testing the local Lorentz invariance foundation of General Relativity}
\label{SRtests}     
\begin{tabular}{p{4cm}p{3cm}p{3cm}}
\hline\noalign{\smallskip}
Experiment & Sector & notable references \\
\noalign{\smallskip}\svhline\noalign{\smallskip}
Cesium atomic clock & matter $(p,n)$ & \cite{wolf2006,Bars:2019lek} \\
Optical clock & matter $e$ & \cite{Sanner:2018atx} \\
Hydrogen maser & matter  & \cite{phillips2000}  \\
$^{29}Ze$ and $^3He$ masers & matter $n$ & \cite{Cane:2003wp} \\
Optical cavity & photon & \cite{Parker:2015ena,Michimura:2013kca} \\
Microwave cavity & photon & \cite{Lipa:2003mh,Stanwix:2005yv,Nagel:2014aga} \\
Spin-polarized pendulum & matter & \cite{Heckel:2006ww} \\
Penning Traps & matter & \cite{BASE:2015mmu} \\
Meson oscillation tests & matter $(K, B, D)$ & \cite{KLOE-2:2013ozx,LHCb:2016vdl} \\
Neutrino experiments & matter $(\nu_e, \nu_\mu, \nu_\ta )$ & \cite{MINOS:2008fnv,IceCube:2017qyp} \\
\noalign{\smallskip}\hline\noalign{\smallskip}
\end{tabular}
\end{table}

\begin{figure}[h]
\sidecaption
\includegraphics[scale=.40]{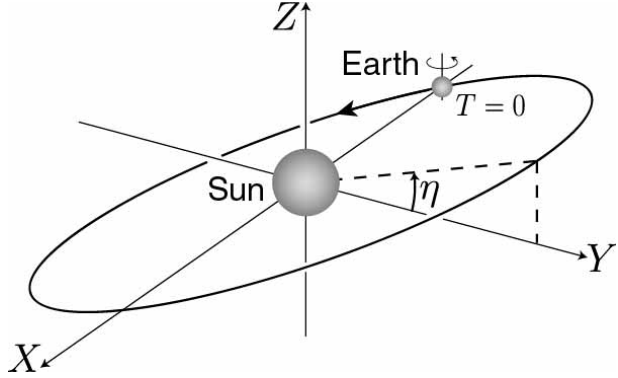}
\caption{Sun-centered Celestial Equatorial System.
Figure used with permission from Ref.\ \cite{bk06}.}
\label{scf}  
\end{figure}

Standard assumptions for experiments is to assume constancy of the coefficients in an adopted standard reference frame of the SCF inertial coordinates \cite{kl99,km02,datatables}.
For the time scales of most experiments 
($\le 100$ years), 
this system is inertial.
The coordinates are usually denoted with capital letters ($T,X,Y,Z$) and the system is depicted in
Figure \ref{scf}.

The coefficients in \rf{photon} have been measured in Earth laboratory experiments involving resonant cavities.
Precision resonant cavity tests in laboratories search for variations in resonant cavity frequency as the Earth rotates and revolved with with respect to the background
coefficients.
Focusing on the ``mass dimension 4" $(k_F)^{\mu\nu\ka\la}$ coefficients, 
the leading effects on the cavity 
frequency shift $\de \nu$ for an optical cavity takes the form
\beq
\bal
\fr {\de \nu}{\nu} &= (-\fr 12 \sin \ch \cos \ch [3 \ep_+ (\tilde{\ka}_{e+})^{YZ} + \ep_- (\tilde{\ka}_{e-})^{YZ}] + ... ) \sin \om T
\\
&+ ( -\fr 14 \sin^2 \ch [3 \ep_+ (\tilde{\ka}_{e+})^{XY} + \ep_- (\tilde{\ka}_{e-})^{XY}] + ... ) \sin 2\om T + ...,
\label{cavity}
\eal
\eeq
where the ellipses includes contributions from a subset of the other coefficients in $k_F$.
Here $\ch$ is the laboratory co-latitude, 
$\om$ is the Earth rotation frequency, 
and $\ep_+, \ep_-$ are dependent on cavity properties \cite{km02}.

Resonant cavity tests,
including the ones in Table \ref{SRtests},
have now measured a subset of $9$ dimensionless coefficients in $(k_F)^{\mu\nu\ka\la}$
to better than parts in $10^{-15}-10^{-19}$.
This indicates of course that
Lorentz invariance holds to an extraordinary degree for photons.
While ultimately this test framework does not provide predictions for the sizes of the coefficients \cite{kp95}, 
one can compare this to a benchmark value of one Planck mass suppression
$m_e /M_{Planck} \approx 10^{-22}$.

It should be noted that one can directly compare measurements of coefficients in these tests.  
For example, 
the ${\tilde c}_{00}=m_p c_{00}$ coefficient for the proton has been measured with a Cesium fountain clock with no signal observed at the $10^{-15}$ level \cite{Pihan-LeBars:2016pjg}.
This can be compared with limits obtained from high-energy cosmic rays
of order $10^{-22}$ \cite{coleman99,km13}.
Note that for the latter limits, 
some assumptions are made about astrophysical properties, 
and so they can be considered not as rigorous as a controlled laboratory experiment.

\subsubsection{Free-fall and gravity measurements}

Precision tests on Earth in the last decades include a plethora of sensitive short-range gravity experiments, 
modern laboratory WEP tests, 
and a vast collection of gravimeter data (including gravimeters based on atom interferometers).

\begin{figure}[h]
\sidecaption
\includegraphics[scale=.50]{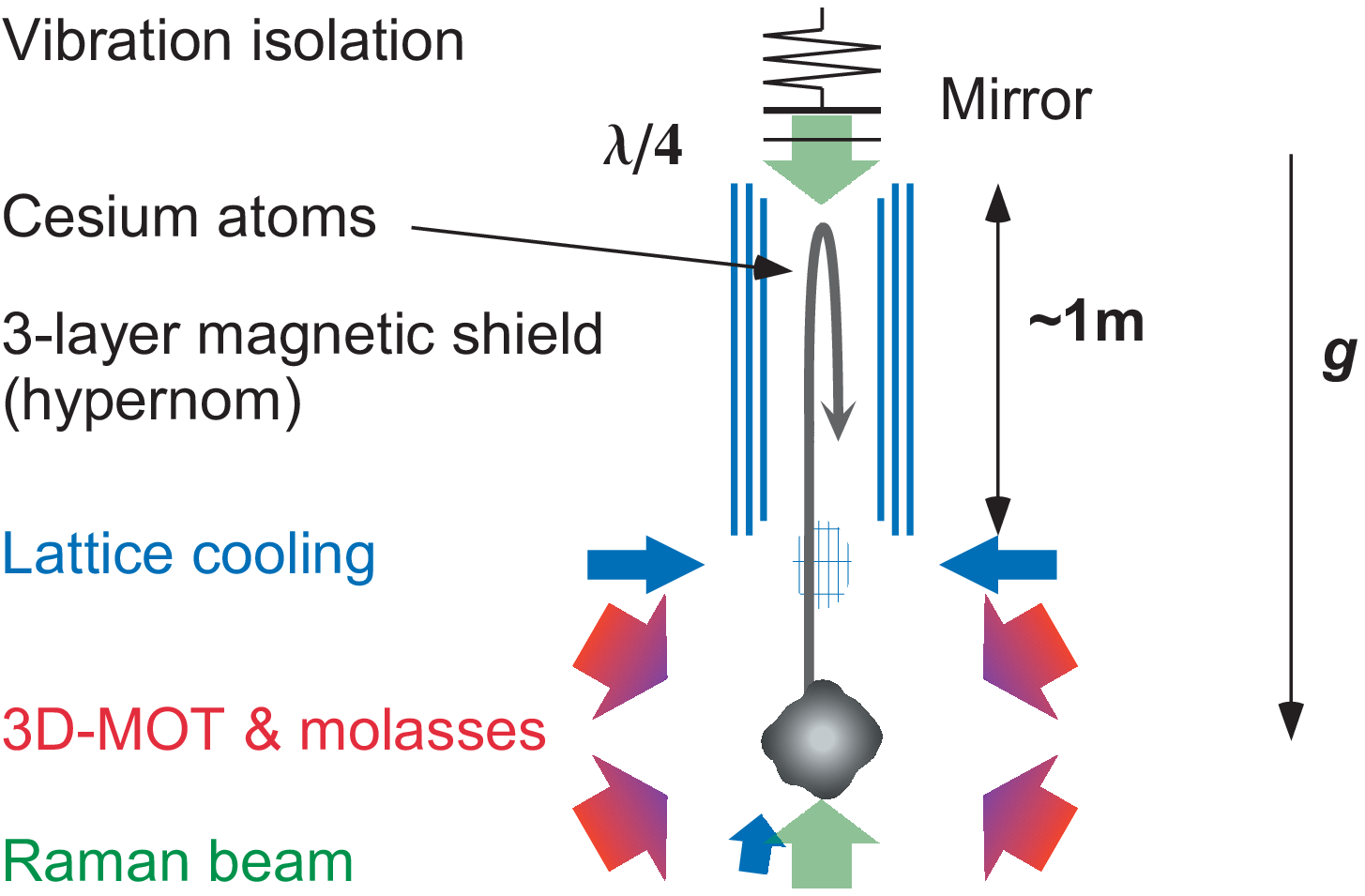}
\caption{The atom interferometer apparatus. Figure used with permission from Ref.\ \cite{Chung:2009rm}. }
\label{AIexp}       
\end{figure}

\begin{figure}[h]
\sidecaption
\includegraphics[scale=.50]{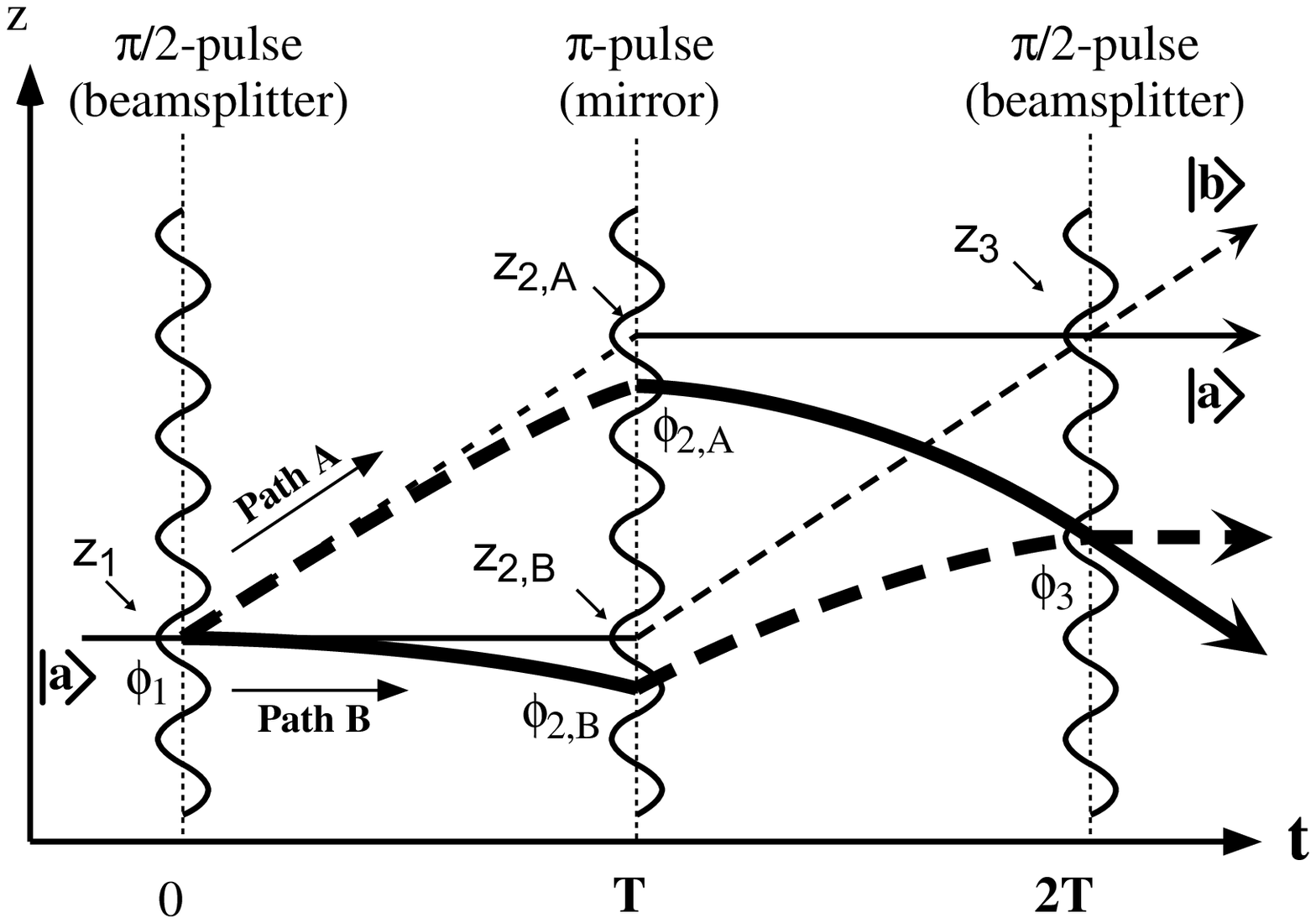}
%
%
\caption{A schematic diagram for the Mach-Zehnder interferometer.  Figure used with permission from Ref.\ \cite{Chung:2009rm}.}
\label{AIexp2}       
\end{figure}

Local measurements of $g$ on Earth with gravimeters can probe for possible WEP and Lorentz violations.
For example, 
one interesting state-of-the-art piece of equipment is the atom interferometer apparatus, 
shown to be able to measure $g$ with high precision \cite{peters99}.
In figure \ref{AIexp2} there is a schematic of the basic principle of the Mach-Zehnder interferometer. 
In figure \ref{AIexp} there is a schematic of the apparatus.

The interferometer can measure the accumulated phase difference between the two paths in \ref{AIexp2}.
In a gravitational field this is $\De \ph = k_j g_j t^2$, 
with $t$ being the time elapsed, 
$g_j$ being the local gravitational field, 
and $k_j$ the momentum transfer in the beamsplitter.
Experiments with atom interferometers worldwide have obtained astounding precision \cite{Merlet_2010,Tino:2020dsl}.
Measurements of coefficients in, for example, 
the result \rf{delg}, 
have been performed.
In the SCF, 
the signal takes the form
\beq
\bal
\fr {\de g}{g} &= \frac 14 (\sb^{XX} - \sb^{YY} ) \sin^2 \ch \cos (2 \om T + 2\ph )  +  \frac 12 \sb^{XY} \sin 2\ch \cos (2 \om T + 2\ph )+ ...
\\
& - \frac 14 V_{\oplus} \sb^{TY} (\cos \et - 1)\sin^2 \ch \cos ( (2 \om +\Om) T + 2\ph ) 
 +...,
 \label{delg2}
\eal
\eeq
where the ellipses includes other terms at other combinations of the Earth's sidereal rotation frequency
$\om$ and orbital frequency $\Om$.  
Here $\ch$ is the laboratory co-latitude and $\et$ is the angle in Figure \ref{scf}.

On the Earth, 
one has to carefully subtract large Newtonian tidal signals already present.
Once this is done, 
one can obtain constraints on $7$ of the {\it a priori} independent coefficients $\sb_\mn$ at the level of in parts in $10^9$ \cite{Muller:2007es}.
Improvement from analysis of a network of worldwide superconducting gravimeters has also been achieved \cite{Shao:2017bgz}.
It should also be pointed out that constraints on $\sb_\mn$ also exist in lunar laser ranging \cite{Bourgoin:2016ynf}
solar system \cite{Hees:2015mga,Poncin-Lafitte:2016nqd} and astrophysical tests \cite{Abbott_2017}
(see a review in \cite{Hees:2016lyw}.

Redshift tests performed since Pound and Rebka are numerous.
See \cite{Will:2018bme} for an exhaustive review.
Most notably, 
near-Earth rocket tests have tested $\al$ in \rf{redshift} to the level of parts in $10^5$ \cite{vessot80,Delva:2018ilu}.
A measurement of the gravitational redshift effect with an atom interferometer on Earth
was reported in Ref.\ \cite{Muller:2010zzb}, 
achieving a sensitivity of $\approx 10^{-8}$ on $\al$,
with some controversy ensuing \cite{Wolf:2010yz}.
Future missions to improve sensitivities include using space-based clock missions like ACES and STEP \cite{Savalle:2019isy,Overduin:2012uk}.

In experiments, 
one can study the effects on two clocks with 3 types of redshift tests. 
This include clocks at the same location moved around in the gravitational field (``a null" type redshift test), 
the classic redshift test comparing
clocks at different locations, 
and synchronizing two clocks and having one move around a closed path (``twin-paradox" type tests).
In fact, 
a work using the different types of redshift and gravimeter tests was used to constrain the EFT coefficients $a_\mu$ for the proton, neutron, and electron \cite{Hohensee:2011wt}.

Modern WEP tests have improved sensitivities over original tests by orders of magnitude.
One of the best recent Earth based tests of WEP used a rotating torsion balance comparing beryllium and titanium \cite{Schlamminger:2007ht}.
The near-Earth satellite test, 
MICROSCOPE, 
has produced the strictest limits on WEP violation
so far \cite{Touboul:2017grn}, 
comparing titanium and platinum test bodies.

Regarding the modified force in \rf{modforce}, 
analysis of gravimeters worldwide can be used to measure WEP/Lorentz violation parameters \cite{Flowers:2016ctv}.
Analysis from MICROSCOPE yields limits on coefficients for Lorentz violation that break WEP \cite{Bars:2019lek}.
The best limits from this experiment on the $a_\mu$ coefficients for the proton, neutron and electron are of order $10^{-7}$ to $10^{-14}$ $GeV$, but they do not completely disentangle the coefficients for these species.
In some scenarios these coefficients could be of order of a fermion mass (e.g. an electron mass of $m_e=10^{-3} GeV$), thus these limits push beyond those levels.

\subsubsection{Short-range gravity experiments}
\label{sr}

In the figure below, 
recent results for the Yukawa parametrization 
\rf{yuk} with different experiments are shown.
Note that different types of short-range tests are designed to probe different regions of the $\al-\la$ parameter space.
These regions can correspond with various model predictions or estimates of 
scales for which new physics could be measured.

\begin{figure}[h]
\sidecaption
\includegraphics[scale=.65]{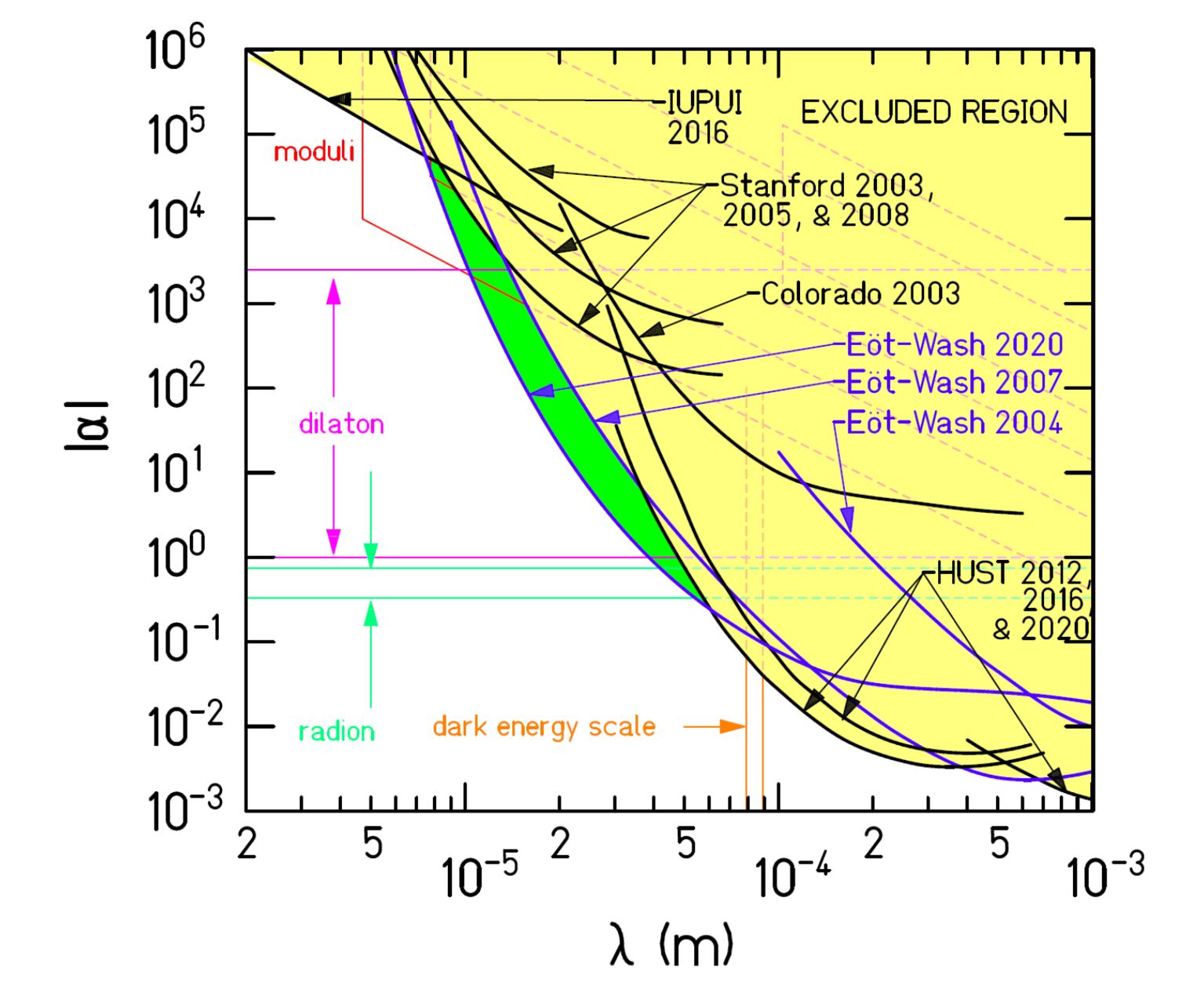}
%
%
\caption{The $\al - \la$ parameter space included the excluded regions in the upper right corner.
Figure used with permission from Ref.\ \cite{Lee20}.}
\label{alphalambda}       
\end{figure}

\begin{figure}[h]
\sidecaption
\includegraphics[scale=.65]{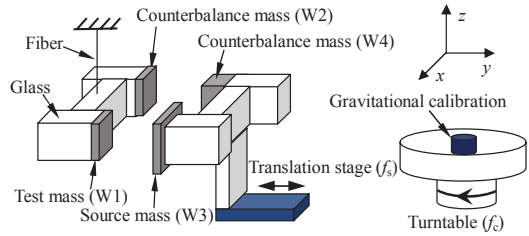}
\caption{The design of the HUST short-range gravity experiment \cite{Shao:2015gua}.  Figure used with permission from the authors of Ref.\ \cite{Shao:2015gua}.}
\label{HUSTapp}       
\end{figure}

As indicated in Figure \ref{alphalambda}, 
there has been a long history of precision short-range gravity tests \cite{Fischbach:1999bc}.
We highlight an example with one of the (Huazhong University of Science and Technology) HUST experiments \cite{Yang:2012zzb}.
This group uses a novel test mass design
which can probe the $\al$ parameter better at longer distances, 
and thus has sensitivity to the Newtonian force.
A sketch of the apparatus is shown in the figure \ref{HUSTapp}.
Limits from this test on the Lorentz violation quartic force terms in \rf{pot} were obtained.
No new signal was found and the limits are of order $10^{-9} \, m^2$ on 14 of the $(k_{eff})_{jklm}$ coefficients \cite{Shao:2015gua}.
Improvements followed when combined with analysis of another experiment \cite{Shao:2016cjk}.

\begin{figure}[h]
\sidecaption
\includegraphics[scale=.30]{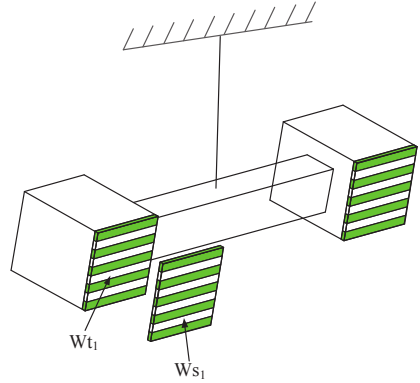}
\caption{A novel striped mass distribution designed to amplify sensitivity to the signal in \rf{pot} \cite{Shao:2016jzh}.
The source mass is $Ws_1$ and the test mass is $Wt_1$.
The figure is used with permission of the authors of Ref.\ \cite{Shao:2016jzh}.}
\label{striped}       
\end{figure}

A sharper sensitivity to the Lorentz violation signal in \rf{pot} can be achieved with a novel
striped design \cite{Shao:2016jzh}.
The latter feature is due to the sensitivity of the signal relying on changes (derivatives) of the matter distribution.  
A rapidly varying mass distribution enhances the effect.
Figure \ref{striped} displays how the source and test mass would be arranged in a torsion pendulum setup.

Other cutting edge experiments are indicated in the figure \ref{alphalambda}.
Notable recent tests include those
by the E\"otwash group \cite{Lee20}, HUST \cite{Yang:2012zzb}, Decca \cite{Decca:2005qz}, and Colorado \cite{Long2003}.
For an exhaustive list and figures of many such tests see the review article of short-range gravity tests in Ref.\ \cite{Murata:2014nra}.

\section{Frontiers}
\label{sec:3}

\subsection{Non-Riemann gravity searches}

There is general interest in beyond Riemann gravity searches like Torsion, 
non-metricity and other non-Riemann gravity scenarios. 
We highlight here some intriguing new spin-dependent and spin-independent gravity couplings in the EFT framework.
The results show a host of effects that could distinguish 
a theory of non-Riemann geometry from one based on Riemann geometry.
Recently, 
in Ref.\ \cite{kl21nR}, 
the phenomenology of the effects of
non-Riemann geometry were calculated in Earth laboratory experiments.
The motivation for this consideration comes primarily from the plethora of recent work on geometries beyond Riemann geometry, 
most notably the Riemann-Finsler geometries 
\cite{Kostelecky:2011qz,Lammerzahl:2012kw,SILVA201474,AlanKostelecky:2012yjr,Javaloyes:2013ika,Schreck:2014hga}.
Basically the idea of a Finsler geometry is that, 
in addition to a spacetime metric $g_\mn$, 
one can define additional structures, 
such as background tensors, 
on a Finsler manifold \cite{matsumoto1986foundations}.
Many of these Finsler geometries can be expected to produce effects analogous to the EFT terms 
discussed below in the action.

An approach to generically identify such terms is to include in the EFT framework, 
terms that appear to break the usual diffeomorphism invariance 
and are not consistent with geometric constraints, 
like $\nabla_\mu G^\mn=0$
\cite{kl21}.
We briefly discuss here some examples of such terms, 
their effect on the acceleration of test particle in the laboratory, 
and specific experiments that would be sensitive to such effects.

At the level of the Lagrange density terms in the EFT framework, 
Ref.\ \cite{kl21} countenances
extra terms (in addition to \rf{Lps}) for a fermion in a weak gravitational field. 
They take the form,
\beq
\bal
\cL &= -(m^{\prime L} )^\mn h_\mn {\overline \ps} \ps - (a^L)^{\ka\mu\nu} h_\mn {\overline \ps} \ga_\ka \ps + ...
\\
&-\frac 12 (c_h^L)^{\ka\mu\nu\rh} h_{\nu\rh}  {\overline \ps} \ga_\ka i \prt_\mu \ps +...
\label{nRlag}
\eal
\eeq
These terms explicitly break diffeomorphism symmetry.
To find the leading experimental signals in Earth-laboratory tests of gravity, 
one calculates the effective nonrelativistic hamiltonian terms using a generalized
Foldy-Wouthuysen transformation (see Refs.\ \cite{kl99,kt11,Xiao18}).

The results of the non-relativistic hamiltonian can be organized into different classes of terms depending on whether they are spin-dependent, 
involve gravity couplings to the local acceleration $\vec g$, 
and involve powers of the momentum of the particle $\vec p$.
The standard result is
\beq
H_0 = \fr {\vec p^2}{2m} - m \vec g \cdot \vec z - \fr {3}{4m} 
[ 
\vec p^2 \vec g \cdot \vec z + \vec g \cdot \vec z \vec p^2  - (\si \times \vec p) \cdot \vec g 
].
\label{H0}
\eeq
This result has been previous discussed in, 
for example, 
Ref.\ \cite{silenko05}.

The contributions from symmetry breaking, 
non-Riemann geometry 
terms, 
take the form of a series of terms with
subscripts $\{ \ph,g,\si,p \}$.
These subscripts indicate the dependence on the gravitational potential $\ph$,
the acceleration $\vec g$, 
the spin via the Pauli matrices $\vec \si$,
and powers of momentum $p^j p^k...$.
The full result in Ref.\ \cite{kl21nR} is lengthy and so we display here samples of each kind of term:
\beq
\bal
H &= (k^{NR}_{\ph}) \vec g \cdot \vec z
+
\frac 12 (k^{NR}_{\ph p})^j ( p^j \vec g \vec z + \vec g \cdot \vec z p^j)+...
\\
&+
(k^{NR}_{\si \ph})^j \si^j \vec g \cdot \vec z +...+ (k_g^{NR})^j g^j + (k_{gp}^{NR})^{jk} p^j g^k+...
\\
&
+ (k_{\si g}^{NR})^{jk} \si^j g^k
+ (k_{\si g p}^{NR})^{jkl} \si^j g^k p^l
+ ...
\label{HLV}
\eal
\eeq
The coefficients are all labeled with $NR$ which indicates 
they are the nonrelativistic combinations of the coefficients in the original Lagrange density \rf{nRlag}.
Novel effects that show up here are terms that depend on the gravitational potential, 
terms that modify the freefall acceleration of matter with and without spin, 
and terms that would modify the gravity-induced phase shift in matter interferometers.

Experiments can probe many of the signals in the Hamiltonian above.
Some analysis has already been performed with Ultra-Cold Neutrons (UCNs) in the qBOUNCE experiment \cite{Ivanov:2021bvk}, 
obtaining measurements, 
like $|(k_\ph^{NR})_n|<10^{-3} \, GeV$, 
indicating that surprisingly large violations of this type could still exist.
Some other experiments sensitive to effects from \rf{HLV} include
spin-polarized torsion pendula \cite{Heckel:2008hw}, 
experiments measuring the acceleration of atoms with and without spin
\cite{Tarallo14}, 
and neutron interferometers \cite{Haan14,parnell20}.

\subsection{Antimatter Tests}

As a continuation of the theme of testing the WEP principle, 
some theorists have countenanced the notion 
that antimatter could fall differently in a gravitational field (see for example, 
\cite{Goldman:1985rq,Chardin:1996qs,Nieto:1991xq} ).
Recent precision control of laboratory experiments producing antimatter
for a significant finite amount of time makes these tests possible.
This makes precise measurements of differing rates of freefall 
for antimatter and matter possible.

As an example, 
consider the experiment of the anti-hydrogen trapping ALPHA experiment
\cite{ALPHA:2013skd}. 
As a figure of merit, 
the best achievement so far is
\beq
\frac {M_g}{M_i} < 75
\label{eta}
\eeq
at the 5\% confidence interval.
Nonetheless, 
the tests remain very challenging.
Other proposed tests include GBAR
\cite{Perez:2015zya}, 
for which dramatic improvements 
have been forecast in simulations \cite{Rousselle:2022yof}.

As a guide to how these effects might arise, 
first consider matter-gravity couplings discussed above in the EFT.
In Ref.\ \cite{kt11} it was shown in a simplified model (a subset of the EFT) that
an anomalous acceleration for antimatter over matter could occur if the coefficients
$\ab_\mu$ and $\cb_\mn$ obeyed a special relation like
\beq
\al (\ab^w)_T = \frac 13 m^w (\cb^w)_{TT},
\label{IPM}
\eeq
where isotropy of the coefficients is assumed in a special coordinate system
and $w$ runs over the species $e,p,n$.
for antimatter, 
the sign of $\ab_\mu$ changes while for $\cb_\mn$ it remains the same.
Thus one can have the inertial and gravitational masses of Hydrogen equal while
for anti-Hydrogen they would not be equal, 
yielding anti-matter falling with a different acceleration.
This model can also be generalized to include nonminimal terms in the EFT \cite{kv15}, 
so that the relation in \rf{IPM} includes a series of coefficients with different mass dimension.
Furthermore, 
the non-Riemann couplings discussed above could impact antimatter freefall experiments, 
in a similar manner \cite{Kostelecky:2021tdf} but with the {\it a priori} independent coefficients
of \rf{nRlag}.

\begin{acknowledgement}
The author thanks Cosimo Bambi, Alan Kosteleck\'y, Zhonghao Li, and Jay Tasson for useful comments and the generous approval of the authors of other published works to re-use many of the figures in this article. 
The author was supported during the completion of this work by the United States National Science Foundation, award number 2207734.
\end{acknowledgement}

\bibliographystyle{myspringer}

\bibliography{refs}

\end{document}